\documentclass[aps,prl,preprint,footinbib]{revtex4}
\usepackage[T1]{fontenc}
\usepackage[small,bf]{caption} 
\usepackage[latin9]{inputenc}
\usepackage{graphicx} 
\usepackage{amssymb}
\usepackage{amsmath}

\usepackage{subfigure} 

\usepackage{color}
\usepackage[normalem]{ulem}

\begin{document} 

\title{Unfolding the Sulcus} 
\author{Evan Hohlfeld} 
\affiliation{Department of Physics, Harvard University, Cambridge, Massachusetts 02138} 
\affiliation{Lawrence Berkeley National Lab, Berkeley, California 94720} 
\author{L. Mahadevan} \affiliation{Engineering and Applied
Sciences,  Harvard University, Cambridge, Massachusetts 02138}

\begin{abstract}
Sulci are localized  furrows on the surface of soft materials that form by a compression-induced instability. We unfold this instability by breaking its natural scale and translation invariance, and compute a limiting bifurcation diagram for sulcfication showing that it is a scale-free, sub-critical {\em nonlinear} instability. In contrast with classical nucleation, sulcification is {\em continuous}, occurs in purely elastic continua and is structurally stable in the limit of vanishing surface energy. During loading, a sulcus  nucleates at a point with an upper critical strain and  an essential singularity in the linearized spectrum. On unloading, it quasi-statically shrinks to a point with a lower critical strain, explained by  breaking of scale symmetry. At intermediate strains the system is linearly stable but nonlinearly unstable with {\em no} energy barrier. Simple experiments confirm the existence of these two critical strains. 
\end{abstract}

\maketitle

\begin{figure} \includegraphics[width=1\linewidth]{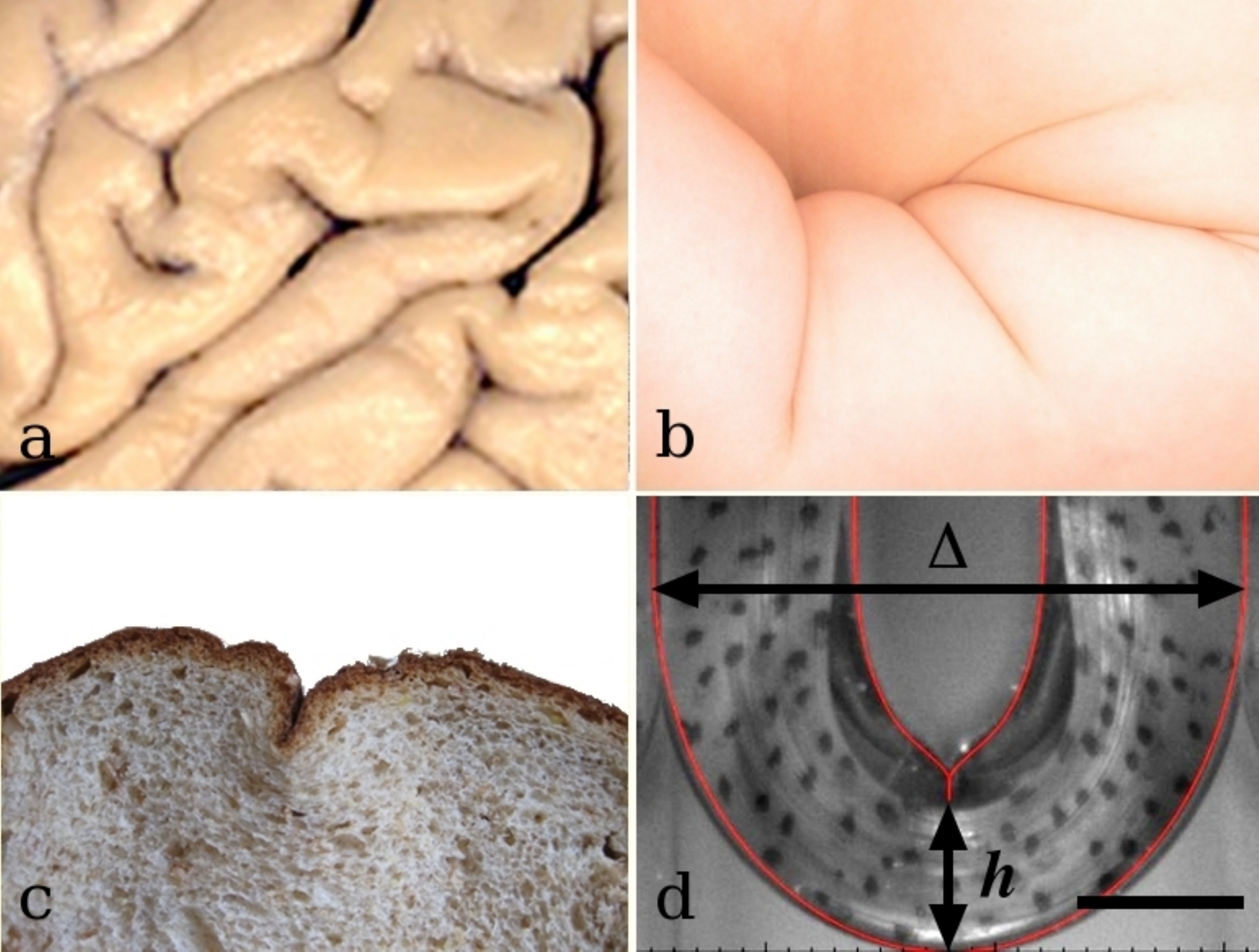} \caption{Examples of sulci in (a) a primate brain \cite{Bradbury}, (b) the arm of an
infant, (c) sliced bread under lateral compression, and (d) the bent slab of PDMS used in our experiments. The results of a numerical simulation shown in
red capture the form of the sulcus in the gel, as described in the text, with no adjustable parameters. The scale bar represents 2.3 cm in (a)  5 cm in (b)  2.5 cm in (c)  and 0.33 cm in (d).}
\label{sf:arm_creases} \end{figure}

Sulci are usually seen in combination with complementary protrusions known as gyri on the surface of the primate brain, but are also seen on the palm of our hand, in our elbows and knees, in swollen cellular foams (such as bread) and gels, and in geological strata; a few representative examples are shown in Fig. \ref{sf:arm_creases}. While observations of sulci are ancient, their systematic study is fairly recent; an early reference is the reticulation patterns in photographic gelatin \cite{Sheppard}, and there has been a small interest in these objects both experimentally \cite{Southern,Tanaka, Hayward,
Gent, Ghatak} and theoretically \cite{Tanaka,Hwa, Onuki89, Silling}, starting with the pioneering work of Biot \cite{Biot} over the past 50 years. Despite
this, there is no careful analysis of the fundamental instability and bifurcation that leads to sulci. Here we study the formation of a sulcus in a
bent slab of soft elastomer, e.g. PDMS: as the slab is bent strongly, it pops while forming a sulcus that is visible in the lower right panel of Fig. \ref{sf:arm_creases}; releasing the bend causes the sulcus to vanish continuously, in sharp contrast with familiar hysteretic instabilities that pop in both directions. We find that sulcification is a fundamentally new kind of nonlinear subcritical surface instability with no scale and a strongly topological character, yet has no energetic barrier relative to an entire manifold of linearly stable solutions.  We also argue that sulcification instabilities are relevant to the stability of soft interfaces generally, and provide one of the first physical examples of the consequences of violating the complementing condition \cite{ADN} (during loading) and quasiconvexity at the boundary \cite{JMB84} (during unloading),  keystones in the mathematical theory of elliptic partial differential equations and the calculus of variations.

To understand the unusual nature of sulcification, we first recall Biot's calculation for the linear instability of the surface of an infinite half-space of an incompressible elastomer that is uniformly stressed laterally. Because the the free surface is the softest part of the system, and there is no characteristic length scale in the equations of elasticity or in the boundary conditions, instability first arises when the Rayleigh surface wave speed vanishes. For incompressible rubber, Biot \cite{Biot} showed that this threshold is reached when the compressive strain exceeds $45.3\%$ \cite{Biot}, at which value, all surface modes are unstable while the fastest growing one has an infinite wave number. Since every free surface looks like a half-space locally, Biot's instability lurks at every free boundary. Finite geometries typically break this infinite degeneracy and lead to a hierarchy of ordinary buckling instabilities that preempt the surface instability \footnote{Thus the Biot instability is actually the progenitor of all structural or buckling instabilities.}. However, since Biot's calculation was limited to a linear analysis of the problem, it could not address the question of whether the instability was supercritical or
subcritical or its ultimate nonlinear saturation. Since the basic problem is scale free and translation invariant (the sulcification instability can arise anywhere along the surface), the nonlinear problem is numerically intractable without explicit regularization and a careful limiting process requiring that we unfold the sulcus literally and figuratively.

Therefore we consider the bent strip geometry shown in Fig. \ref{sf:arm_creases}d, and  break scale invariance by assuming that a thin skin of a stiff material is attached to the surface of the bent slab. Furthermore, the curvature maximum at the bottom of the horseshoe where the highest strains are achieved naturally breaks translation invariance. For planar deformations, the total energy of the system is given by \[E\left(\mathbf{x}\right)=\frac{\mu_{0}}{2}\int_{\Omega}\left(\left|\mathbf{\nabla x}\right|^{2}-2\right)d^{2}\mathbf{X}+\frac{B_{0}}{2}\int_{\Gamma}\left| \frac{d^{2}\mathbf{x}}{ds^{2}}\right|^{2}ds.\] where $\mathbf{x}\left(\mathbf{X}\right)$ is the deformation of a strip occupying a
rectangular material volume $\Omega\subset \mathbb{R}^{2}$ and subject to the incompressibility constraint $\det\left(\mathbf{\nabla x}\right)=1$, $\mu_{0}$ is the shear modulus of the incompressible elastomer, $B_{0}$ is the stiffness of a semi-flexible skin and $s$ is the arc length parameter of the upper surface $\Gamma\subset\partial\Omega$. Our model corresponds to having a simple neo-Hookean elastomer free energy for the bulk and a Bernoulli-Euler curvature energy for the skin. To understand the onset of the sulcification instability, we extremized the energy above using a custom-built finite element method with continuous strains and a hierarchical mesh (see Supplementary Information (SI)). We enforced incompressibility and self-contact using pressure fields and by assuming left-right symmetry about the sulcus. This model has three relevant length scales: a regularization length $l_{r}=\sqrt[3] {\frac{B_{0}}{\mu_{0}}}$, the length of the strip $L_{s}$, and its thickness $W_s$. We use $\mu_{0}$ and $L_s$ to scale all quantities so that $B=\frac{B_{0}}{\mu_{0}L_{s}^{3}}$ and the aspect ratio of the strip $L_s/W_s$ are the only dimensionless parameters.

\begin{figure}
\subfigure[]{\includegraphics[width=.6\linewidth]{figure2a}\label{sf:bifurcation-diagram}}
\subfigure[]{\includegraphics[width=.3\columnwidth]{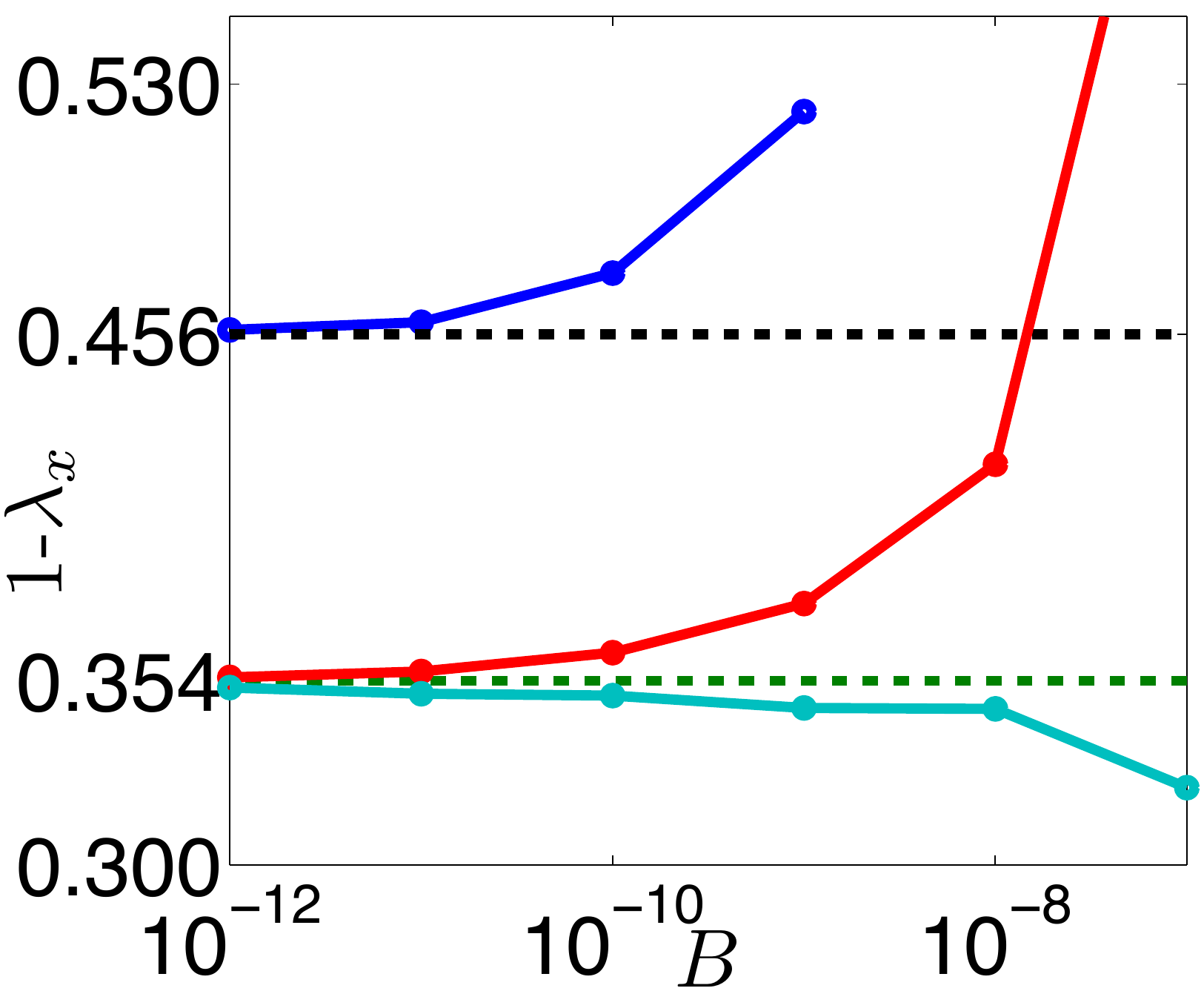}\label{sf:strain-convergence}}\subfigure[]
{\includegraphics[width=.3\columnwidth]{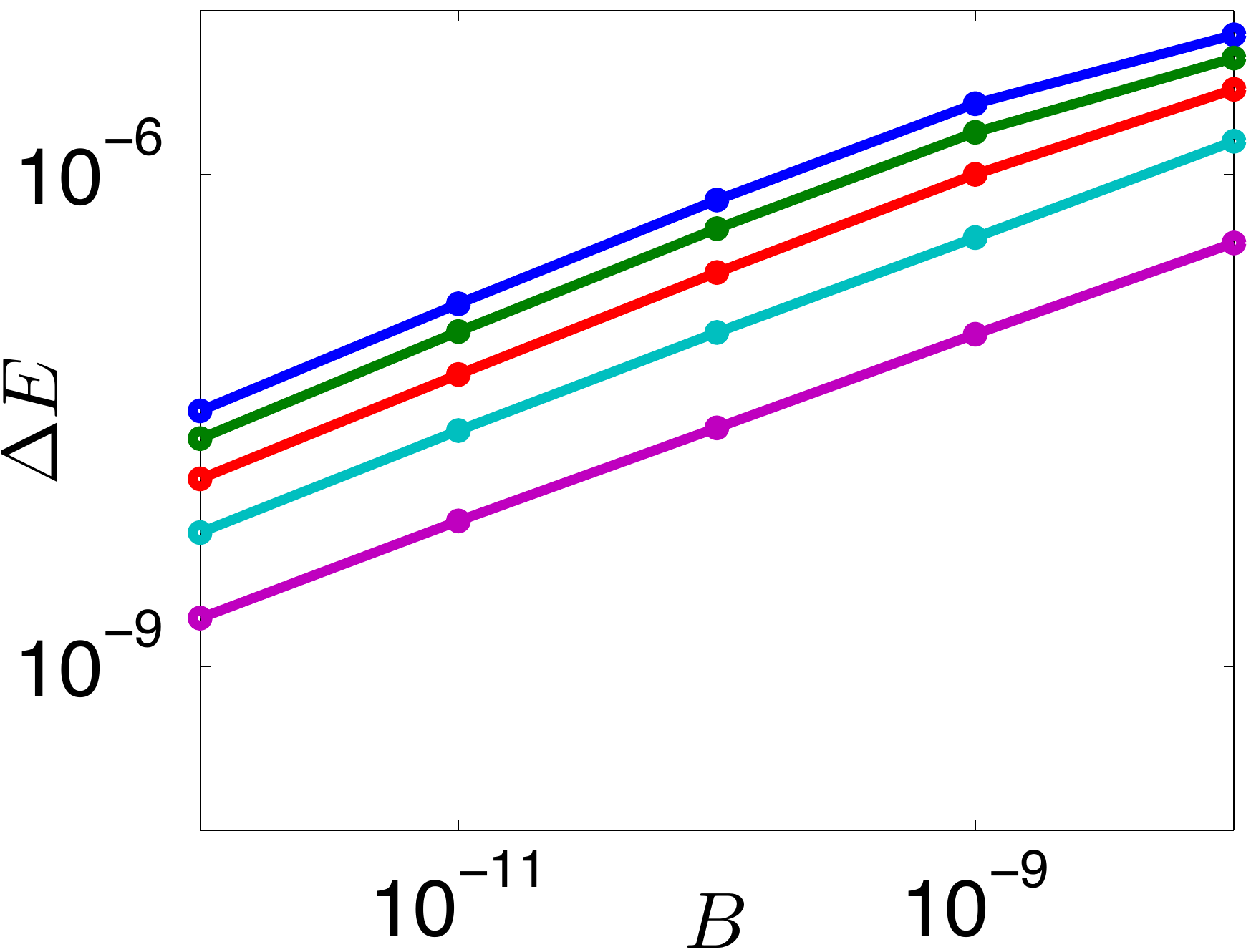}\label{sf:energy-barrier-convergence}}
\caption{ (a) Bifurcation diagrams for the bent strip geometry shown in Fig. 1 (d), showing the scaled height $h$ as a function of the scaled compression $\Delta$, the bifurcation parameter for $B\in \left[10^{-7}-10^{-12}\right]$ (yellow-magenta); solid lines mark stable equilibria and dotted lines mark unstable equilibria; thick gray lines highlight the asymptotic T-shaped diagram delimited by the Biot point on the left and the T point on the right. Solutions on the upper stable branches have smooth surfaces. Insets a-c show the structure (color=strain energy) for typical solutions on the other branches, rescaled to a fixed size.  (b) The upper two curves show compressive strain in terms of the principal stretch $\lambda_{x}$ at the eventual location of the sulcus $\mathbf{X}=\mathbf{0}$ on the smooth branch for values of $\Delta$ at the right and left fold points, respectively, for each $B$. The lowest curve shows the smallest value $1-\lambda_{x}$ attained on the surface outside the sulcus for the solution at the right fold point. The dashed lines correspond to the upper-critical strain predicted by Biot (upper line), and the new lower-critical strain extrapolation from our simulations (lower line). (c) The energy barrier $\Delta E$ for sulcification as a function of $B$ sampling the solution at even intervals between the  left (bottom curve) and right fold (top curve) points.} 
\end{figure}

Our simulations start with an initially flat rubber strip that is bent and quasistatically compressed between parallel, rigid plates separated by a distance $\Delta$. As the control parameter $\Delta$ is decreased, compressive strain on the inner surface of the strip increases and ultimately drives sulcification. To ensure that the scale of the furrow is not numerically under-resolved, we use a recursively refined finite element mesh near the incipient sulcus to keep the mesh scale roughly an order of magnitude smaller than $l_{r}$ (See SI, Sec. B). Using a novel continuation method for variational inequalities (see SI, Sec. C), we computed both stable and unstable extrema of $E\left(\mathbf{x}\right)$, and explored the limit $B \to0$. This yields the central result of our study, the family of bifurcation diagrams shown in Fig.\ref{sf:bifurcation-diagram}, where we plot the minimum height of the slab $h$ as a function of $\Delta$.

Each $h-\Delta$ curve is a bifurcation diagram for a different value of $B$: solid lines represent linearly stable solutions, while the dotted lines
represent linearly unstable solutions. Each curve follows the characteristic S-shape of a hysteretic transition, associated with a sudden change in $h$ and formation or relaxation of a finite size sulcus when a critical value of $ \Delta$ is passed in loading or unloading. For every value of $B$, extrema on the top branch have smooth surfaces, those on the middle branch have a pendant of size $l_r$ [see inset a of Fig. \ref{sf:bifurcation-diagram}], and those on the bottom branch have a self-contacting sulcus (insets b and c). As $B$ is decreased, ($B \in [10^{-7}-10^{-12}]$, $l_{r}/L_{s} \in [4.6\times10^{-3}- 10^{-4}]$), the hysteresis in a typical loading cycle becomes atypically one-sided as the branch of unstable extrema (dotted lines) swings up toward the top stable branch and the S-shaped bifurcation diagrams converge to a master T-shaped diagram traced by the thick gray line with two critical points.

For fixed $\Delta$ and $B$ an unstable solution represents a saddle point in the energy landscape; its energy relative to the configuration with a flat
surface is $\Delta E=E_{u}\left(\Delta\right)-E_{s}\left(\Delta\right)$ where $E_{u}\left(\Delta\right)$ and $E_{s}\left(\Delta\right)$ are the energies of the unstable and top, stable branches at $\Delta$ respectively, is an upper bound on the height of the barrier to nucleating a sulcus. Fig. \ref{sf:energy-barrier-convergence} shows $\Delta E$ as a function of $B$ and confirms the convergence of the family of bifurcation diagrams toward the limiting T-shaped bifurcation diagram, as well as the existence of a nonlinear surface instability with no energetic barrier over an extended range of
strains. We see that the instability thus differs significantly from traditional first order 
phase transitions in that the deformation is continuous, occurs in simple elastic continua and is well defined in the limit of vanishing surface energy. The presence of metastable region bracketed by a pair of critical strains along which the stable and unstable solutions 
coincide  as the skin becomes vanishingly thin naturally explains the discrepancy between Biot's prediction and a large
number of experiments on creasing and sulcification \cite{Gent} (and references therein) over the past half century. Unfolding
the instability without breaking translation symmetry at the surface, e.g. in a swollen, adhered layer of gel, then naturally leads to extreme sensitivity to imperfections, and  a hierarchy of complex subcritical instabilities connecting Biot's instability and buckling (see SI, sec. E), and the ability to control sulcification \cite{Hayward}.

As $B \rightarrow 0$, the sequence of saddle-node fold points encountered during loading converges to a limiting, infinitely sharp fold point when the
surface strain at the lowest point on the inner surface of the horseshoe, $\mathbf{X}=\mathbf{0}$, reaches a critical value of $45.6\%$ consistent with
Biot's classical result; in Fig. \ref{sf:strain-convergence} we see the convergence of the critical compressive stain, $1-\lambda_{x}$ where
$\lambda_{x}$ is the principal stretch of the deformation gradient $\mathbf{\nabla x}$ along the free surface, for finite $B$ to Biot's predicted value at $B=0$. A numerical linearized spectral analysis of the loaded slab also confirmed Biot's prediction that the  Rayleigh surface wave speed vanishes at $\mathbf{X}=\mathbf{0}$ just as the critical strain is achieved, and corresponds to the failure of the complementing condition \cite{Hohlfeld:2008fk, ADN}, wherein infinitesimal periodic solutions at the boundary grow at a rate that diverges as the inverse of the wavelength. Over-damped dynamical simulations--which trace steepest descent contours of the energy landscape--reveal that nonlinear effects reorganize these surface waves into a self-similar furrowing process; after a short transient, depending on $B$, the growth of a sulcus, which occurs via rolling, not snapping, is described by the 
self-similar form $ \mathbf{x}_{s}\left(\mathbf{X},\Delta^{\star}\right)+\sqrt{\lambda t}\mathbf{v}^{\star} \left(\mathbf{X}/\sqrt{\lambda t}\right)$ where $\lambda$ is a dimensional constant and $\mathbf{v}^{\star}$ is the numerically computed scale-invariant form of the sulcus \cite{Hohlfeld:2008fk}, and $\mathbf{x}_{s}\left(\mathbf{X},\Delta\right)$ is the branch of smooth solutions and  $\Delta^{\star}$ is the value of $\Delta$ at Biot's limiting fold point.

The complementary sequence of saddle-node fold points for decreasing $B$ encountered during unloading are actually ``corners'' associated with the loss/gain of a self-contacting sulcus as the unstable surface pendant just closes to form a cavity of fixed size $l_{r}$. As these corners converge to
the limiting ``T-point'', the maximal surface strain outside the self-contacting region approaches a limiting value of $35.4\%$ that is attained at a
sequence of points converging to $\mathbf{X}= \mathbf{0}$. The convergence to this strain is traced by the lowest curve in Fig. \ref{sf:strain-convergence} with the asymptote marked by the dashed line. The middle solid curve of Fig. \ref{sf:strain-convergence} is another estimate of the critical strain
computed by measuring the strain at $\mathbf{X}=\mathbf{0}$ for a sequence of extrema for corresponding values of $\Delta$ on the top branch. The T-point critical strain (like the Biot critical strain) is universal for free surfaces of incompressible materials, consistent with recent experimental observations \cite{Gent, Hayward}; however they both change with applied normal stress (i.e. indentation), for compressible materials \cite{Hohlfeld:2008fk} etc.

To understand why the T-point bifurcation and the entire unstable manifold are not captured by linearized analysis, we note that before the sulcus reaches the regularization scale $l_{r}$,  it shrinks according to the form $\mathbf{x}_{s}\left(\mathbf{X},\Delta\right)+l\left(\Delta\right)
\mathbf{v}_{T}\left(\mathbf{X}/l\left(\Delta \right)\right)$ where $l\left(\Delta\right)\ge0$ vanishes at the T-point and $\mathbf{v}_{T}\approx\mathbf{v}^{\star}$ (See insets $b$ and $c$ in Fig. \ref{sf:bifurcation-diagram}, and the relative scale factor of $6.5$.) Since the elastic stress is determined by $\nabla\mathbf{x}$, this transformation shrinks the size of the sulcus without altering the local stress balance; therefore all the material and contact non-linearities remain relevant even for vanishingly small sulci.  

\begin{figure} \includegraphics[width=1.\linewidth]{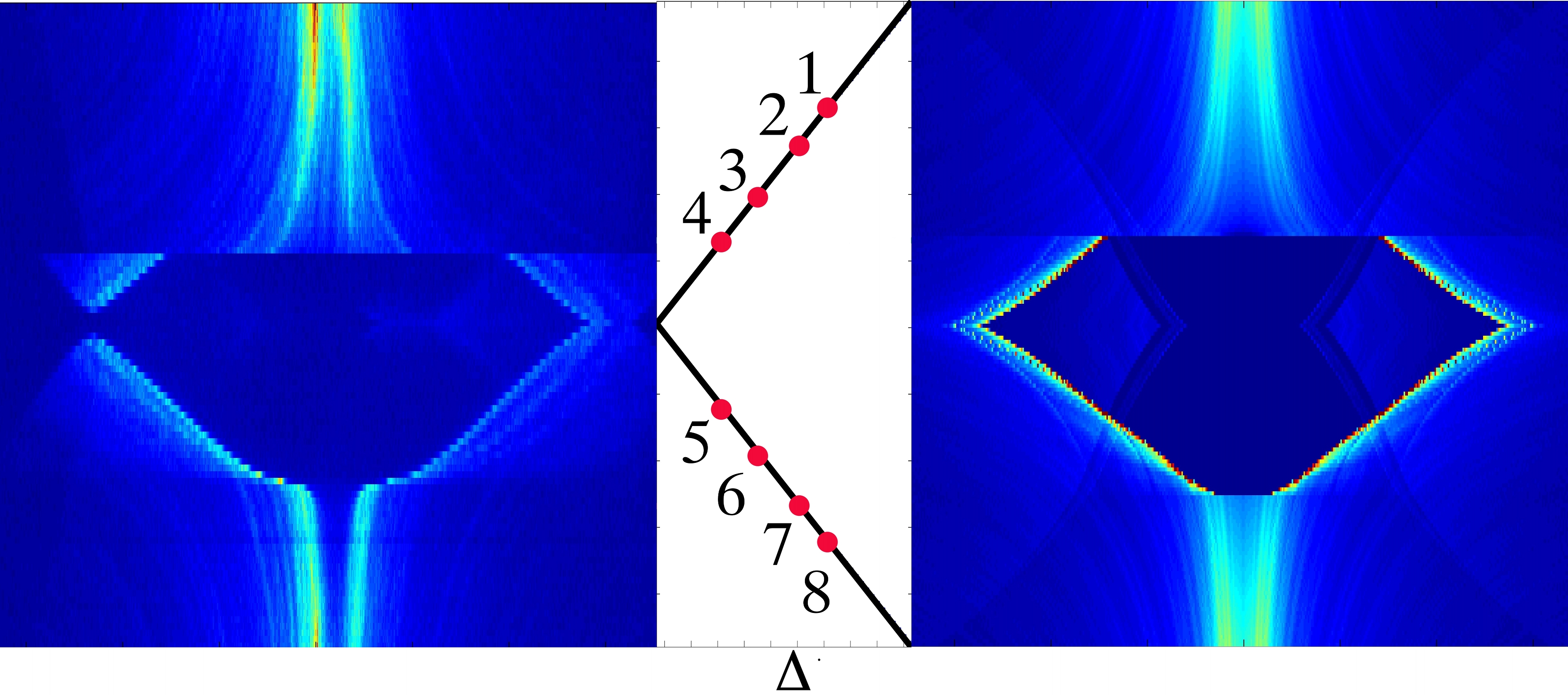} \caption{A comparison of experimental (left) and computed (right) light intensity
patterns cast by the central section of a laser sheet illuminating the slab's bending axis; the horizontal axis is transverse to the sheet and the vertical
axis is the step number in a loading-unloading cycle,  and corresponds to changes in $\Delta$ as shown in Fig. \ref{sf:bifurcation-diagram}; 4 = formation of sulcus, when a caustic suddenly forms,  7 = vanishing of sulcus}
\label{sf:caustic-history-theory} \end{figure}

We tested our theory with experiments using a commercial Sylgard 180 Elastomer to form $36\times 26\times 4\, mm$ slabs that were placed between parallel rigid plates attached to linear motors and compressed in small increments of $200\,\mu m$ in a second, separated by $50 s$ to allow for the equilibration of the slab. We tracked sulcification optically by imaging the refracted image of a laser sheet that passed through the slab along its bending axis. When the sulcus formed it sharply refocused the laser sheet into an almond-shaped caustic pattern surrounding a dark shadow (SI, sec. D). Fig.
\ref{sf:caustic-history-theory} (left) shows the evolution of a central raster scan of the caustic pattern during a loading cycle (vertical axis) (see
SI, sec. D). Analogous ray traced light distributions for the simulation, using the measured laser profile and assuming left-right symmetry, the measured system geometry, and $B=10^{-11}$ (physically $l_{r}\approx18\,\mu m$), are also shown in Fig. \ref{sf:caustic-history-theory}(right) for comparison; the numbered red dots correspond to the numbered red dots in Fig. \ref{sf:bifurcation-diagram}. We see that with no free parameters we can capture the one-way hysteretic transition associated with sulcification.

The  emergence of the $T$ and Biot points, and intervening metastable region in the $B\to 0$, can be understood in terms of a {\em nonlinear} generalization of Biot's half-space problem. All our simulations show that when a  half-space of incompressible elastomer is compressed by   $34.5\%$ it has an infinite degeneracy of energy minimizers: the trivial flat configuration, and a continuous family of isolated sulci which are stable up to translation and rescaling  (i.e. $\mathbf{v}\left(\mathbf{X}\right)\to l\mathbf{v}\left(\mathbf{X}/l\right)$ for any number $l>0$), i.e., these symmetries are spontaneously broken.  Sulcification exchanges compressive strain for rotation and shear which are ultimately limited by self-contact.  Beyond the lower critical strain, forming a sulcus of size $O(l)\ll 1$ releases energy over a region of size $O(l^2)$, equivalent to the failure of quasiconvexity at the boundary. The spatial variations of $\mathbf{\nabla x}$  near the compressive strain maximum at $\mathbf{X}=\mathbf{0}$ act as symmetry-breaking perturbations and determine the ultimate scale $l$ of the surface fold. When the compressive strain is not localized to a point, the size of the sulcus is not set by the local geometry of $\mathbf{\nabla x}$ and the  domain, and  the $T$ bifurcation is sensitive to details and potential interactions between multiple sulci resulting in reticulation \cite{Sheppard, Tanaka}, or in a combination of buckling and sulcification \cite{Cerda, Boudaoud} etc. (see SI, sec. E).

More generally, $T$ bifurcations might arise in elastic systems with internal interfaces and  nucleationlike processes in elliptic systems where nonlinearities enter in a scale-free way, e.g. the formation of cavities, bubbles and cracks \cite{Gent59,JMB82}. These processes are notoriously difficult to control, displaying extreme sensitivity to imperfections, and are associated with a discontinuous transition in the microscopic state characterized by a critical size nucleus; e.g, a bubble or crack will grow only once it has reached a threshold size. $T$ points should exist in these systems in the limit when the surface energy vanishes and the size of the ``defect'' also vanishes, but the ratio of the two which corresponds to a critical pressure or stress remains finite. 

Acknowledgments: We thank D. Cuvelier for help with the experiments and the Kavli Institute and the Harvard MRSEC for support.
\bibliographystyle{plain} 
 
\newpage
\appendix
\begin{center} {\Large Supporting Information for ``Unfolding the sulcus'' \\ by E. Hohlfeld and L. Mahadevan} \end{center}

Here we provide more details about our simulation techniques, the full form of the inequalities governing the self-contact problem inside 
the sulcus, our finite element discretization, the details of the rough continuation technique and experimental methods together with further experimental results, and a sketch
of some generalizations to multiple buckling, sulcification and dynamics.

\subsection{Equations }

In this first subsection we present, in detail, the equations we have
used to model sulcification. The equations governing the interior
of the rubber block, in dimensionless form, are \begin{align*}
\frac{\partial}{\partial X_{j}}\left\{ S_{ij}+p\mathrm{cof}_{ij}\left(\frac{\partial\mathbf{x}}{\partial\mathbf{X}}\right)\right\}  & =0\\
\det\left(\frac{\partial\mathbf{x}}{\partial\mathbf{X}}\right) & =1\\
S_{ij}=\frac{\partial W}{\partial A_{ij}} & =A_{ij}\end{align*}
where $\mathbf{S}$ is the nominal stress tensor, $p$ is the pressure,
and $W=\frac{1}{2}tr\left(\mathbf{A}^{T}\mathbf{A}-2\right)$ is the
strain energy density used in the main text; $A_{ij}:=\partial x_{i}/\partial X_{j}$.

The contact problem for the surface within the fold of the sulcus
is governed by a system of variational inequalities. We have assumed
left-right symmetry for simplicity. Using a coordinate system where
the $1$-direction is parallel to the initially flat upper surface
of the strip and the $2$-direction is the outward normal to this surface, on
the right half of the upper surface the inequalities describing contact are \[
x_{1}\ge0,\quad\pi\ge0,\quad x_{1}\pi=0.\]
 The first inequality expresses the restriction of the upper surface
to the right half plane -- by symmetry, this is the same as forbidding
self-penetration. The second inequality is dual to the first and restricts
the contact pressure, $\pi,$ to be positive. The final equality is the Karush-Kuhn-Tucker
condition which states that the contact pressure is zero if the upper
surface is not in contact. A similar system of inequalities governs
the interaction of bottom surface of the bent strip with the {}``apparatus''
that bends the strip to drive sulcification on the upper surface.
The regularizing bending stiffness term appears in the boundary conditions
for the upper surface as \[
S_{i2}=B\frac{\partial^{4}x_{i}}{\partial X_{1}^{4}}.\]

The effects of the skin become important to the shape of the sulcus when the sulcus reaches the size $\sim B^{\frac{1}{3}}$. Using the above formula and the scaling form of the sulcus at the T-point we see that \[S_{i2}\sim B\times B^{\frac{1}{3}}\times B^{-\frac{4}{3}}\sim const.\] That is, a critical stress must emerge in the limit that the surface energy and ``defect'' size simultaneously go to zero as explained in the text, but the T-point stress/strain need not be the limiting stress computed here which is a function of position and the form of the surface energy.

\subsection{Finite element discretization.}

To simulate sulcification in the bent strip geometry, we discretized the
equations of the incompressible neo-Hookean model using the finite
element method. (See e.g. \cite{SCB02} for a detailed description
the finite element method.) An example mesh for the finest scale simulations
presented in main text is shown in Supplementary Fig. \ref{fig:The-mesh}. The
elements in the mesh alternatively represent displacement degrees
of freedom as tensor products of Hermite polynomials or pressure degrees
of freedom as quadratic Lagrange polynomials. This combination allows
for continuous strains and stresses, which is important since the
sulcification and Biot instabilities have characteristic strains;
Hermite displacement elements are also required for compatibility
with the surface bending energy. The contact pressure was described
by continuous quadratic Lagrange surface elements. Because left-right
symmetry was assumed, only the right half of the mesh shown was used,
resulting in approximately 9000 degrees of freedom for the finest
scale simulation in the main text.

\begin{figure}
\includegraphics{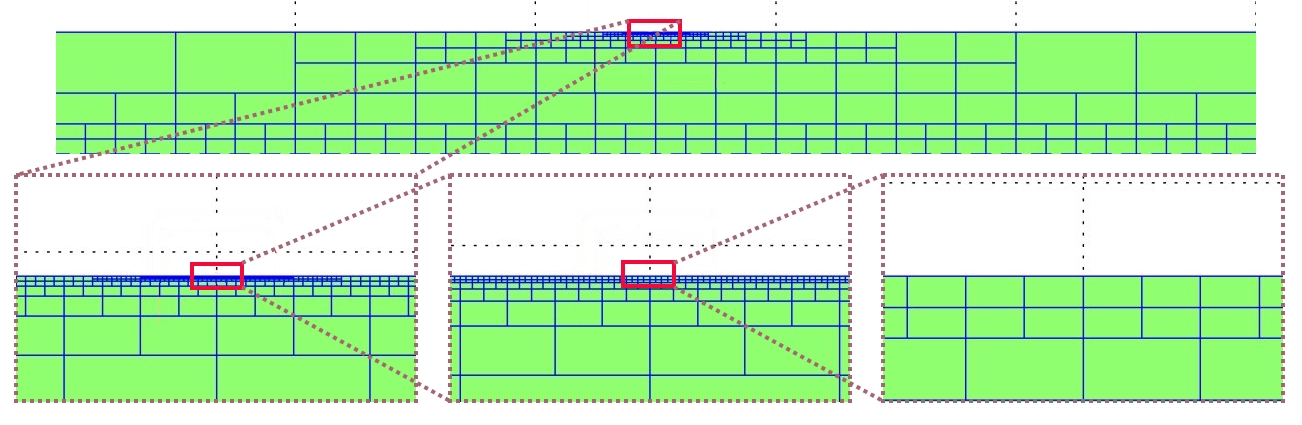}

\caption{The mesh used for the simulations in the main
body of the paper is recursively refined on the surface near the sulcus.
Displacement variables have continuous derivatives and are modeled
as tensor product Hermite elements. Pressure variables are continuous
and modeled as quadratic Lagrange elements.}
\label{fig:The-mesh}
\end{figure}

\subsection{Rough continuation method}

We solved the discretized equations using rough continuation, which
is a variant of pseudo-arclength continuation adapted for variational inequalities.
Pseudo-arclength continuation is described in \cite{ELA80}, it is
Newton's method applied to a system with a control parameter in which
the control parameter is treated as an additional variable on equal
footing with the state variables. Before explaining rough continuation, we will first abstractly describe the usual pseudo-arclength continuation method. 

Let the state of some system be described by the vector $\mathbf{x}\in\mathbb{R}^{n}$, let $\lambda$ be a control parameter, and suppose the system is governed by the equation \begin{equation}
\mathbf{f}\left(\mathbf{x},\lambda\right)=0,\label{eq:equation}\end{equation} where $\mathbf{f}\in\mathbb{R}^{n}$.
 At fixed $\lambda$, a Newton's method iteration is defined as \[
\mathbf{x}_{k+1}=\mathbf{x}_{k}-D\mathbf{f}\left(\mathbf{x}_{k},\lambda\right)^{-1}\left(\mathbf{f}\left(\mathbf{x}_{k},\lambda\right)\right)\]
 where $D\mathbf{f}$ is the $n\times n$ derivative matrix of $\mathbf{f}$.
Pseudo-arclength continuation  computes the connected component of solutions to \eqref{eq:equation} whose closure includes some initial point $\left(\mathbf{x}_{0},\lambda_{0}\right)$ rather that  just a single solution for some specific value of $\lambda$.
An iteration in this method is \begin{align*}
\mathbf{x}_{k+1} & =\mathbf{x}_{k}-P_{k}L_{k}\left(\mathbf{f}\left(\mathbf{x}_{k},\lambda_{k}\right)\right)\\
\lambda_{k+1} & =\lambda_{k}+\delta_{k}\end{align*}
 where $L_{k}$ is any left inverse of the $n\times\left(n+1\right)$
matrix \[
\mathbf{A}_{k}=\left[D\mathbf{f}\left(\mathbf{x}_{k},\lambda_{k}\right),\frac{\partial\mathbf{f}}{\partial\lambda}\left(\mathbf{x}_{k},\lambda_{k}\right)\right],\]
 $P_{k}$ is the projection operator \[
P_{k}=\mathbb{I}-\mathbf{t}_{k}^{T}\mathbf{t}_{k}\]
 in which $\mathbf{t}_{k}\in\mathbb{R}^{n+1}$ is the unit row vector
orthogonal to all the rows $\mathbf{A}_{k}$, and $\delta_{k}$ is
the $\lambda$-component of $\mathbf{t}_{k}$. At the end of $N$
iterations at step $r$, we make a new initial guess for $\left(\mathbf{x}_{1}^{r+1},\lambda_{1}^{r+1}\right)$
by \[
\left(\mathbf{x}_{1}^{r+1},\lambda_{1}^{r+1}\right)=\left(\mathbf{x}_{N}^{r},\lambda_{N}^{r}\right)+\Delta\mathbf{t}_{N}^{r}\]
 where $\Delta$ is a small, usually adaptively chosen step size.
Each iteration of the method accomplishes an approximate projection
back to the solution curve $\mathbf{x}\left(\lambda\right)$. The
method has an ambiguity in the sign of $\Delta$. This is fixed by
setting $\mathbf{t}_{N}^{r}\cdot\mathbf{t}_{1}^{r+1}>0$. Pseudo-arclength
continuation inherits the quadratic rate of convergence of Newton's
method whenever the matrix $\mathbf{A}\left(\mathbf{x},\lambda\right)$
is continuous in $\mathbf{x}$ and $\lambda$.

Both Newton's method and pseudo-arclength continuation can easily
handle constraints posed as Lagrange multipliers, such as the constant
volume constraint of our incompressible strip. Inequality constraints,
like contact, pose a problem, however. Linear inequality constraints
can be posed as \[
\left(\mathbf{C}\cdot\left(\mathbf{x}-\mathbf{R}\right)\right)_{i}\ge0,\quad\left(\mathbf{p}\right)_{i}\ge0,\quad\left(\mathbf{C}\cdot\left(\mathbf{x}-\mathbf{R}\right)\right)_{i}\left(\mathbf{p}\right)_{i}=0\]
 for a fixed vector $\mathbf{R}$ and matrix $\mathbf{C}$ where $\mathbf{p}$
is the vector of Lagrange multipliers, and the notation $\left(\mathbf{p}\right)_{i}$
means the $i^{th}$ component. The active set method can be used to
solve these inequalities at fixed $\lambda$. In this method, a set
of {}``active'' constraints are guessed and the inequality constraints
in the active set are replaced with equality constraints while the
remaining inequality constraints are ignored. A Newton iteration is
then attempted. If the result of the Newton iteration violates other
constraints, these are turned on for the next iteration, while if
a Lagrange multiplier becomes negative, the associated constraint
is turned off in the next iteration. 

Rough continuation is a synthesis of the pseudo-arclength continuation
method with the active set method that solves two interrelated problems.
First, whenever a constraint turns on or off, the matrix $\mathbf{A}$
must change discontinuously. And second, the sign of $\Delta$ is
again ambiguous since each new Lagrange multiplier changes the sign
of $\det D\mathbf{f}$. In practice, these are only problems at parameter
values where a constraint is just turning on or off along the length
of the curve $\mathbf{x}\left(\lambda\right)$, which results in a
discretization induced corner in the otherwise smooth curve $\mathbf{x}\left(\lambda\right)$.
We can solve both problems by simply replacing Newton iterations with
active set iterations in the pseudo-arclength method except when the
set of active constraints is not converging even while the iterations
remain bounded. In this case we attempt to guess the correct {}``forward''
direction on the underlying smooth curve (i.e. without discretization),
and then add a small multiple of $\mathbf{t}_{k}$ to $\left(\mathbf{x}_{k},\lambda_{k}\right)$
at each iteration so to {}``nudge'' $\left(\mathbf{x}_{k+1},\lambda_{k+1}\right)$
in this direction. Hence the resulting projection step is not always
orthogonal to the current tangent $\mathbf{t}_{k}$, but rather orthogonal
to our best guess for the chord connecting $\left(\mathbf{x}_{N}^{r},\lambda_{N}^{r}\right)$
and $\left(\mathbf{x}_{1}^{r+1},\lambda_{1}^{r+1}\right)$. This modification
drives the iterations away from where a constraint is just turning
on or off so that the active set is fixed and ordinary pseudo-arclength
iterations can resume. In practice the method easily handles large
steps in which several constraints turn on or off at once.

\subsection{Experiments}

This subsection presents details of the experimental data presented in the main text and additional results from other experiments that help to characterize sulcification. These additional results include observations of the caustic pattern cast by a fully three dimensional sulcus, observations of the dynamic formation of a sulcus, and measurements of the force applied to the confining plates that bend the PDMS slab into a horseshoe shape and drive sulcification.

\begin{figure}
\subfigure[]{\includegraphics[width=0.3\linewidth]{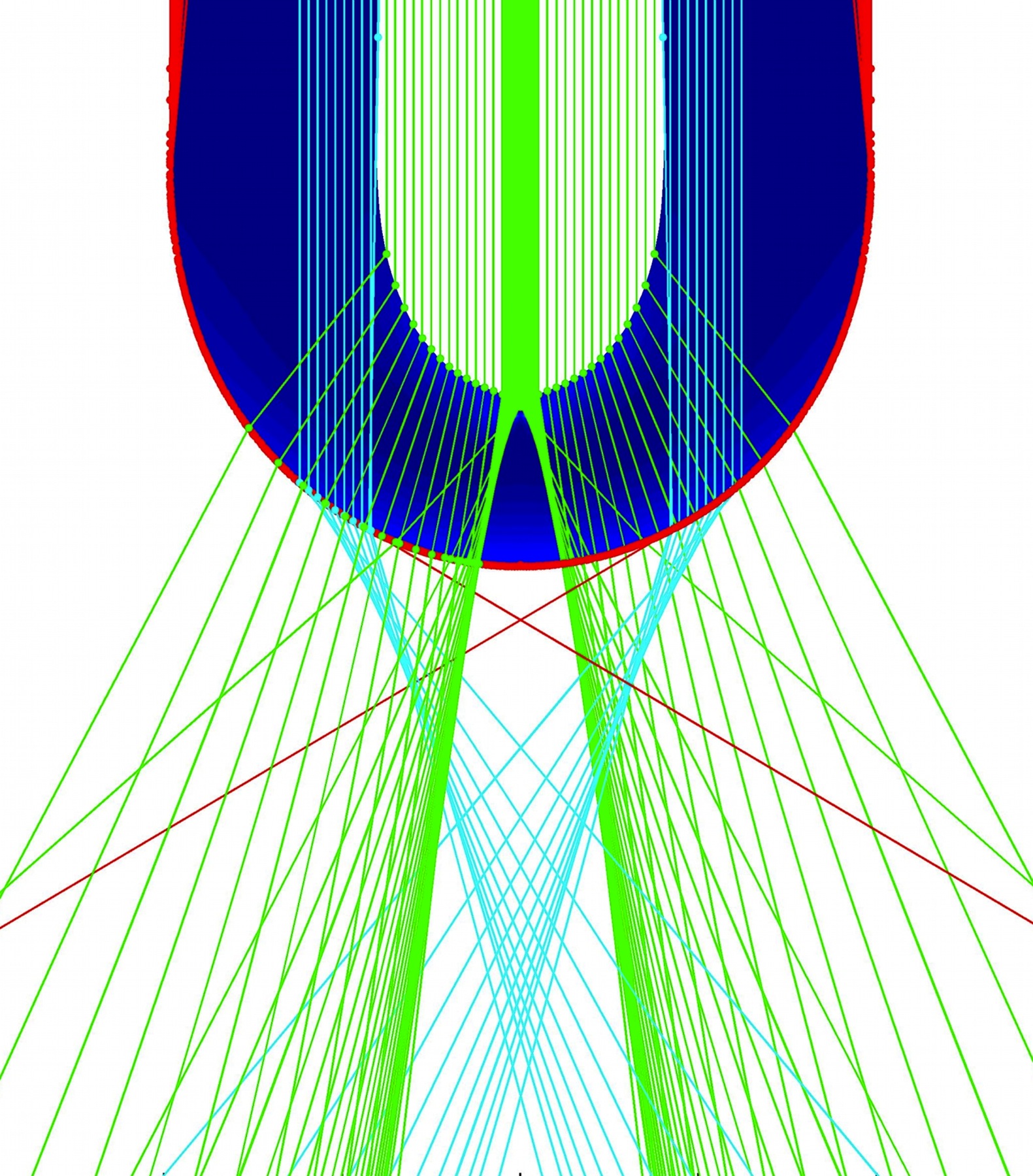}\label{sf:ray-tracing}}
\subfigure[]{\includegraphics[width=0.45\linewidth]{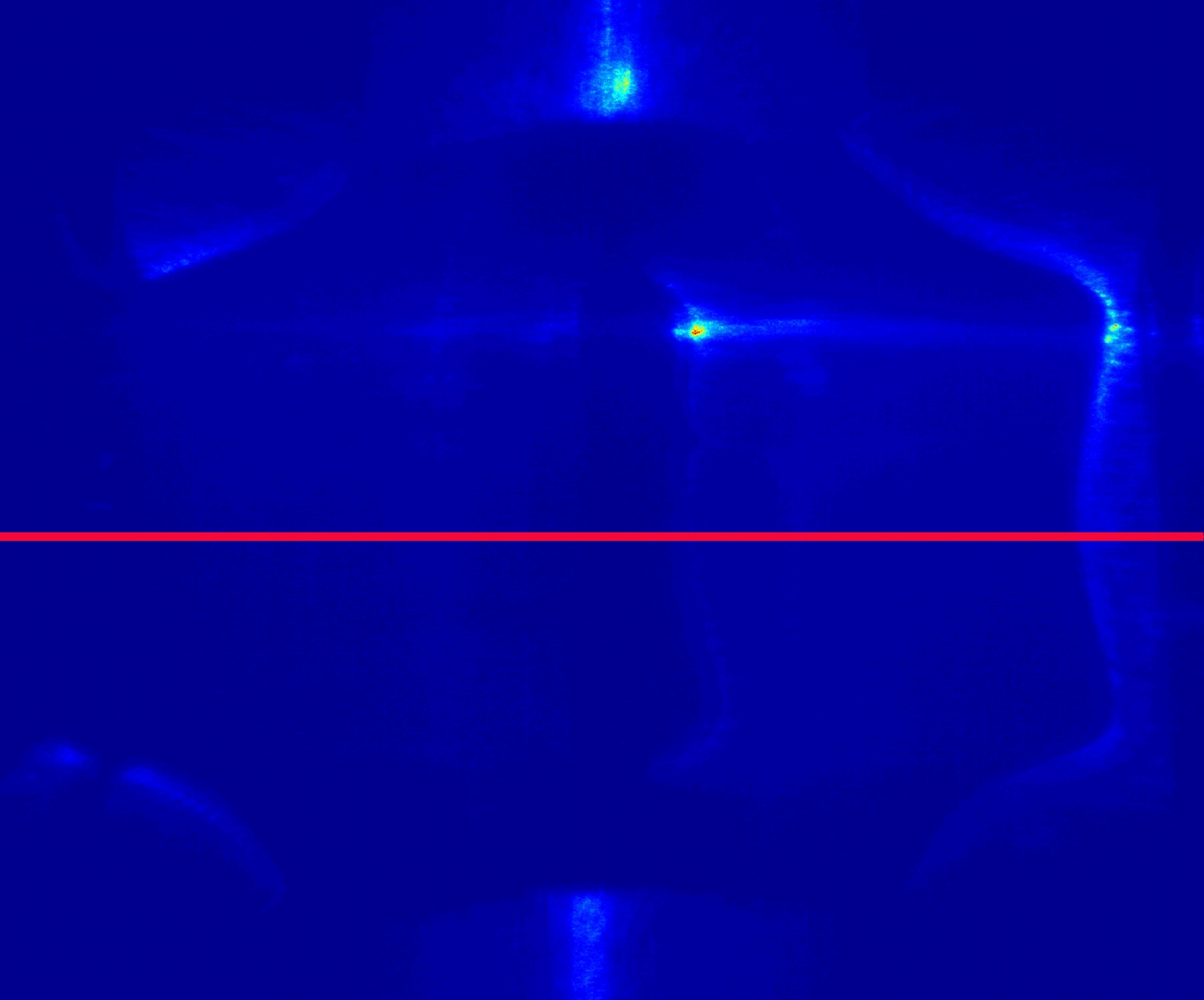}\label{sf:exp-caustic-image}}
\subfigure[]{\includegraphics[width=0.25\linewidth]{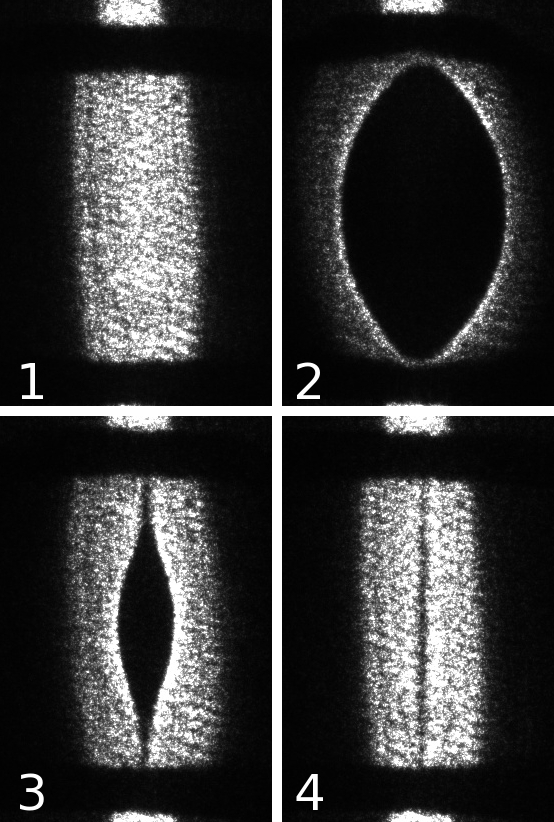}\label{sf:four-patterns}}
\subfigure[]{\includegraphics[width=0.5\linewidth]{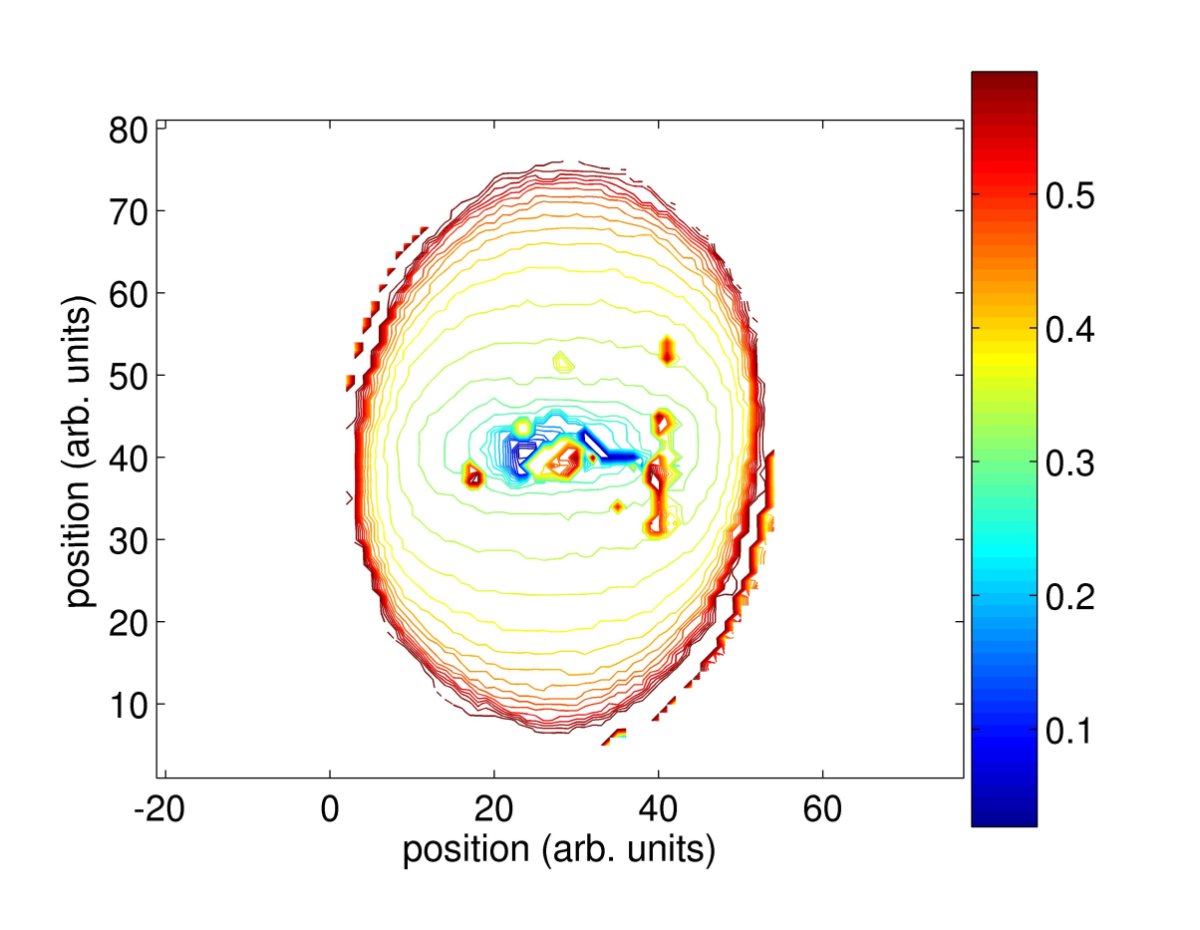}\label{sf:growing-caustic-pattern}}

\caption{Fig. \ref{sf:ray-tracing} shows an example of the ray tracing
used to compute Fig. 4(right) of the main text. In Fig. \ref{sf:exp-caustic-image}
a sulcus at maximum compression refracts a laser sheet into a caustic pattern; the bright
bands at the top and bottom are the unrefracted laser sheet that missed
the sample. The red line is the raster along which data were taken
make Fig. 4(left) of the main text. Fig. \ref{sf:four-patterns} shows refracted light from an another sample at four points in a compression cycle: (1) the initial configuration, (2) the first appearance of a sulcus, (3) as the sulcus quasistatically shrinks to a point, (4) the initial configuration after one cycle. The residual scar seen in (3) and (4) is also visible by eye at glancing angle and vanishes over a period of hours to days. Fig. \ref{sf:growing-caustic-pattern}
shows the evolving shape of the dark shadow cast by developing sulcus
at $20\, ms$ intervals (computed by thresholding frames from a high
speed movie); color indicates time in seconds and the vertical axis
is along the sulcus. In contrast with the quasi-static light distributions,
a detailed three dimensional simulation is needed to calibrate these
data, but we see that the sulcus rapidly grows from a point in a few
tenths of a second.}
\end{figure}

As stated in the main text, we compared the predicted shape of the
sulcus to the experimental shape by making detailed a comparison between
refracted light patterns from laser illumination. Supplementary Fig.
\ref{sf:ray-tracing} shows ray tracing for a bent slab at maximum
compression. The green lines passing through the cylindrically
curved inner surface of the horseshoe are simply refracted as if by
a cylindrical lens, while the green lines passing through the sulcus are refracted
into a dominant caustic pattern surrounding a dark shadow. Light passing through the ends of the horseshoe
is refracted in to a secondary pattern of caustics (blue lines) and
also into a set of whispering gallery modes (red lines). 

A corresponding light pattern for a three dimensional sulcus in the
experimental geometry is shown in Supplementary Fig. \ref{sf:exp-caustic-image}.
The sample is illuminated by a laser sheet along its symmetry axis
(vertical axis in the figure). The unrefracted light which missed
the sample can be seen as the bright blue bands in top and bottom
extremities of the figure. The red line in the figure is the raster
along which data was sampled at intervals in the compression cycle
to make the Fig. 4 (left) of the main text. The outermost bright blue
ring in Supplementary Fig. \ref{sf:exp-caustic-image} corresponds
to the green caustic pattern in Supplementary Fig. \ref{sf:ray-tracing},
the left half of which was partially obscured by the experimental
apparatus. The inner most bright blue ring in Supplementary Fig. \ref{sf:exp-caustic-image}
corresponds to the blue caustic pattern in Supplementary Fig. \ref{sf:ray-tracing}.
The ring-like shape of the caustic patterns is a consequence of edge and three dimensional effects which cannot be described by our two dimensional model.

Supplementary Fig. \ref{sf:four-patterns} shows representative observations of the dark shadow cast by a sulcus in another sample during a loading cycle. In frame (1), the cylindrical curvature of the bent PDMS strip broadened the laser sheet passing through the sample; the unrefracted sheet can be seen as the bright band at the top of each frame, and the dark gap immediately below this band is an artifact of the sample's edge. When the system achieved the critical strain for sulcification, a system spanning sulcus appeared and cast the almond shape shadow seen in frame (2). The shape of the shadow reflects that the sulcus tapered at its upper and lower extrema. Frame (3) shows the shadow pattern shrank in width and depth during unloading and suggests that the sulcus retracted along its length while simultaneously becoming more shallow. Frame (4) shows the initial configuration for the second compression cycle. The clearly visible dark line running down the center of the image suggests that self-adhesion may have prevented the sulcus from completely unfolding; this line is also partially visible in frame (3). We observed that the line gradually filled in over a period hours to days. When the line was present in the refracted light pattern, we also saw a faint line in the surface of an unfolded sample when we observed the sample at glancing incidence.

\begin{figure}
\includegraphics[width=.9\linewidth]{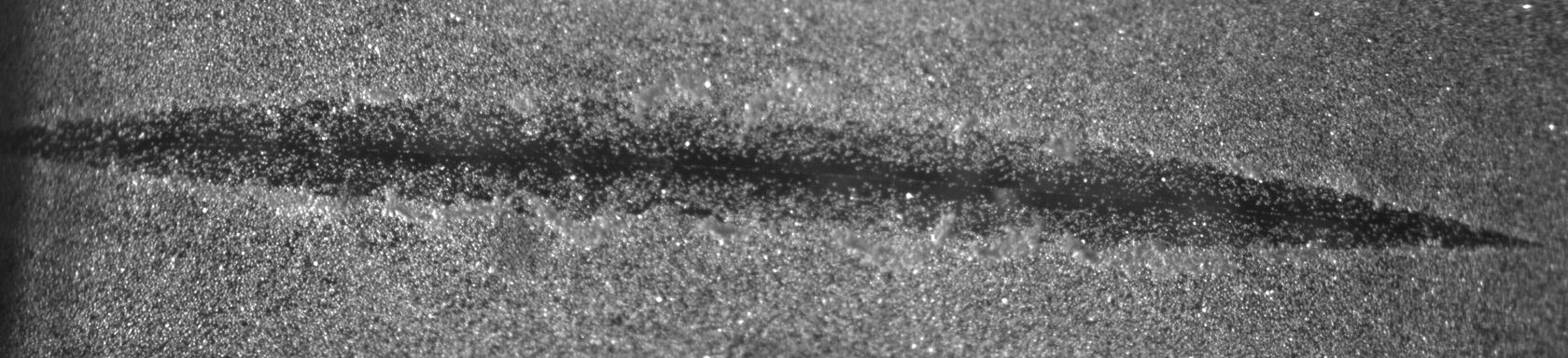}

\caption{Small glass beads dusted on another sample are scattered by the formation of sulcus revealing the self-contacting region when the sample is unfolded as the dark scar crossing the image. The depth of the sulcus changes gradually, tending to zero near the sample's edges, consistent with the idea that the local strain state controls the scale of the sulcus.}
\label{fig:beads}
\end{figure}

Supplementary Fig. \ref{fig:beads} shows how sulcification disturbs a layer of small glass beads dusted on the surface of another sample. The rolling motion of the forming sulcus plows the beads out of the self-contacting region, causing them to pile up in the opening of the furrow. When the sample is unfolded, the depth of the sulcus is made apparent by the width of the depleted region. We see that the depth of the sulcus changes gradually along its length, tending to zero at the sample's edges. This is consistent with the idea that the local strain state sets the (local) scale of the sulcus.

Using high speed video, we observed that the sulcus and corresponding caustic pattern nucleated
at point near the center of bent strip and then grew  to span the width of the bent strip. We tried to quantify this
nucleation process by analyzing high speed movies of the refracted
light pattern, but the detailed patterns and growth speeds were highly
variable. A typical result is shown in Supplementary Fig. \ref{sf:growing-caustic-pattern},
in which the concentric circular lines trace the outline of an observed
dark shadow in the refracted light field at $20\, ms$ intervals;
color also indicates elapsed time in seconds. The vertical axis is
along the length of the sulcus.

\begin{figure}
\subfigure[]{\includegraphics[width=0.45\linewidth]{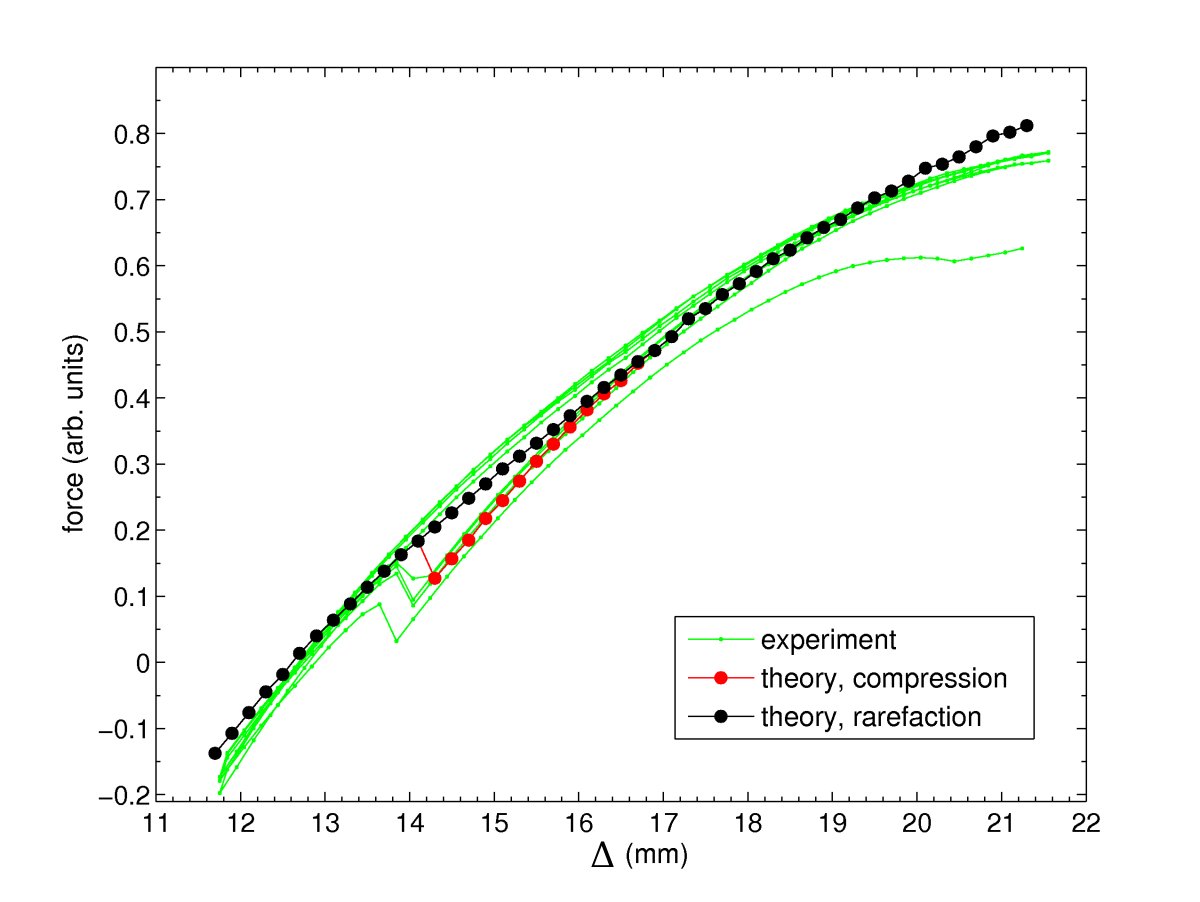}\label{sf:fd-curves}}
\subfigure[]{\includegraphics[width=0.45\linewidth]{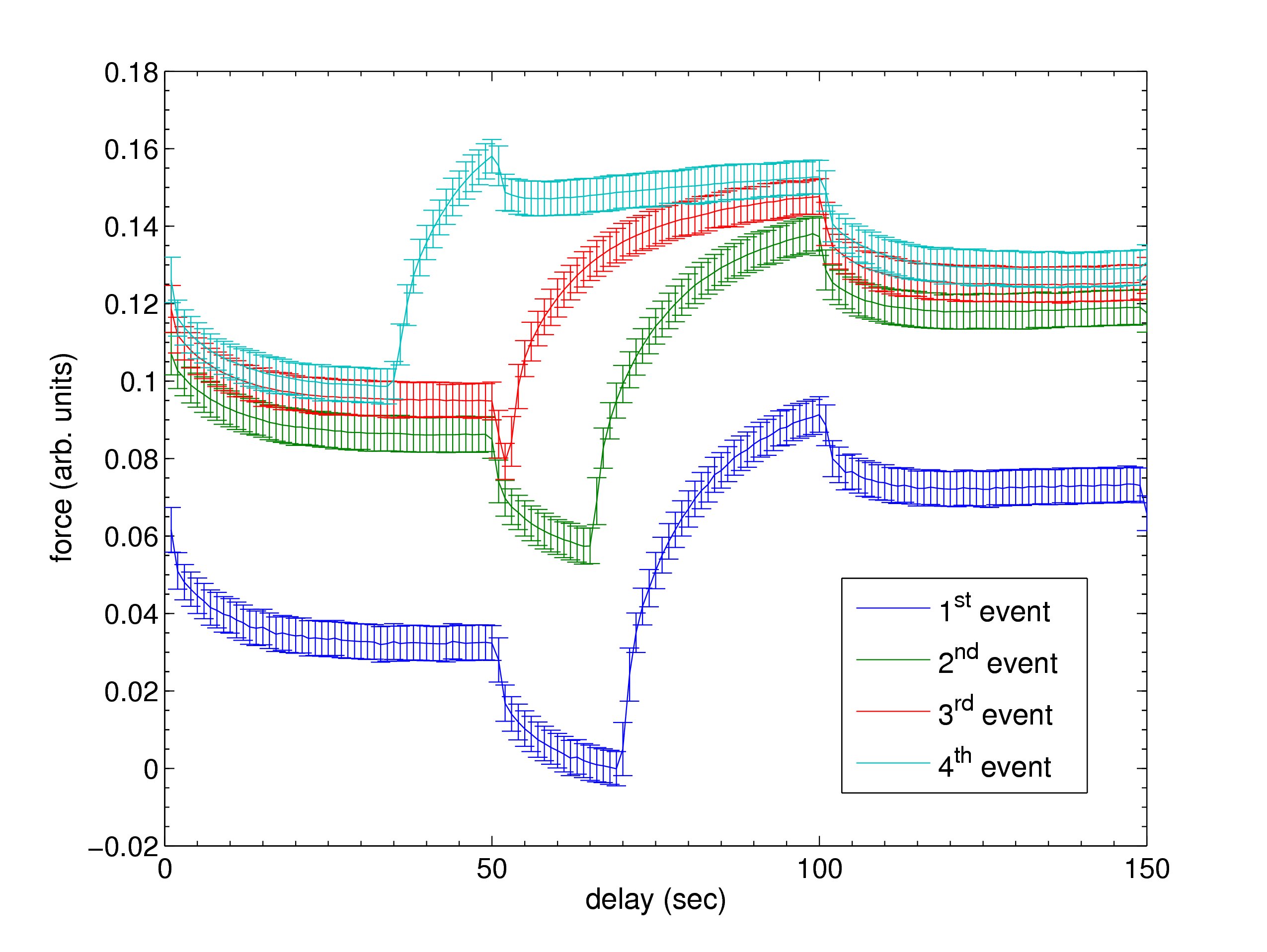}\label{sf:fd-jumps}}

\caption{The green lines in Fig. \ref{sf:fd-curves} show the measured force-displacement curves for multiple loading cycles in the experiment discussed in the main text. The red and black lines are computed from the simulation with the vertical offset and scale of the compression branch (red) as fitting parameters. Fig. \ref{sf:fd-jumps} shows time traces of the  force applied by the compressing plates during the compressive phase of the loading cycles: downward steps coincide with increments to $\Delta$, upward steps indicate sulcification events.}
\end{figure}

Supplementary Figs. \ref{sf:fd-curves} and \ref{sf:fd-jumps} show equilibrium and time series measurements of the force applied to the pair of plates laterally confining the experimental sample. In Supplementary Fig. \ref{sf:fd-curves}, force measurements traced the lower branches of the green lines during compression until a sulcus formed and the force jumped to a less compressive value; the upper green curves were traced during unloading and we did not observe a second  jump expected for typical hysteretic processes. Black and red lines are computed from simulations for $B=10^{-11}$, or $l_{r}\approx18\, \mu m$. We fitted the vertical scale and offset for the red curve to the corresponding section of the green curves. Supplementary Fig. \ref{sf:fd-jumps} shows the time series of force data for the jumps in the green curves of Supplementary Fig. \ref{sf:fd-curves} on successive compression cycles. Downward motion is an increment to $\Delta$, upward motion is a sulcification event. We sampled the force reading at $100\, Hz$ and then averaged in one second windows to produce the curves shown in the figure; the error bars are the standard deviation of the force in an averaging window. There is no apparent correlation between the plate motion (change in $\Delta$) and the precise moment of sulcification. This is consistent with the local nature of the sulcification instability. However, each successive event occurs at lower applied load; this may result from imperfect unfolding of the sulcus due to self-adhesion.

\subsection{Generalizations: multiple creases, buckling and dynamics}

We have argued that nucleation processes, especially in hyperelastic solids with non-convex strain energies, can be understood via the existence of a scale-symmetry breaking solution to a scale-free auxiliary problem which is a nonlinear generalization of Biot's linear problem. In the context of nonlinear elasticity, this problem is to minimize the energy
\[
E_{\mathbf{A},\mathbf{Y}}\left(\mathbf{v}\right)=\int_{F_\mathbf{Y}}W\left(\mathbf{A}+\mathbf{\nabla v}\right)-W\left(\mathbf{A}\right)d\mathbf{X}
\]

over all functions $\mathbf{v}\in W^{1,p}\left(F_\mathbf{Y},\mathbb{R}^d\right)$ where $\mathbf{A}$  is identified with the value of $\mathbf{\nabla x}$ at some point $\mathbf{Y}\in\overline{\Omega}$, $F_\mathbf{Y}$ is all of space if $\mathbf{Y}$ is an interior point and $F_\mathbf{Y}$ is a half-space if $\mathbf{Y}$ is a point on the (smooth) boundary of $\Omega$, $p$ is an exponent related to the growth of $W$ at large $\mathbf{A}$, and $d$ is the dimension of space. Because of the scale-free nature of this auxiliary problem, i.e. $E_{\mathbf{A},\mathbf{Y}}$ is a homogenous function of degree $d$ in $L$ under the mapping 
\[
\mathbf{v}\left(\mathbf{X}\right)\to L\mathbf{v}\left(\mathbf{X}/L\right),\quad L>0,
\]
a minimizer can only exist if
\[
\int_{F_\mathbf{Y}\cap B_r\left(\tilde{\mathbf{X}}\right)}W\left(\mathbf{A}+\mathbf{\nabla v}\right)-W\left(\mathbf{A}\right)d\mathbf{X}\ge 0,\quad\forall\mathbf{v}\in C^1_0\left(B_r\left(\tilde{\mathbf{X}}\right),\mathbb{R}^d\right),\,r>0,\,\tilde{\mathbf{X}}\in F_{\mathbf{Y}}.
\]
This latter condition embodies both quasiconvexity (when $\text{dist}\left(\tilde{\mathbf{X}},\partial F_{\mathbf{Y}}\right)>r$)  and quasiconvexity at the boundary (otherwise). Quasiconvexity was invented by Morrey and is a necessary condition for the functional $E_{\mathbf{A},\mathbf{Y}}$  to be sequentially weakly lower semicontinuous on the Sobolev space $W^{1,p}\left(F_{\mathbf{A},\mathbf{Y}},\mathbb{R}^d\right),$ it is also sufficient  for this property when $W$  is continuous and has polynomial growth in $\mathbf{A}$ \cite{CBM52,EANF84}. Sequential weak lower semicontinuity can be used to prove a minimizer exists when $E_{\mathbf{A},\mathbf{Y}}$ is bounded from below. Quasiconvexity at the boundary was invented by Ball and Marsden as a necessary condition for a local minimizer in the topology of $C^1\left(F_{\mathbf{A},\mathbf{Y}},\mathbb{R}^d\right)$  to be a local minimizer in the comparatively weaker topology of $C^0\left(F_{\mathbf{A},\mathbf{Y}},\mathbb{R}^d\right)$ \cite{JMB84}. Here it is needed to ensure that $E_{\mathbf{A},\mathbf{Y}}$ is bounded from below.

The relationship between quasiconvexity and quasiconvexity at the boundary for our auxiliary problem is analogous to the relationship between ellipticity (in the interior) and satisfaction of the Complementing Condition (i.e. ellipticity at the boundary) for quadratic energy functionals. These conditions result from linearizing the two quasiconvexity conditions at $\mathbf{v}=0$.

The existence of a non-trivial minimizer (or minimizing sequence) for $E_{\mathbf{A},\mathbf{Y}}$ determines the critical deformation gradient $\mathbf{A}$ for nonlinear instability in the original problem and implies that one of the quasiconvexity inequalities is saturated. When a non-zero minimizer exists, say $\tilde{\mathbf{v}}$, there is a continuous family of such minimizers (related by scaling and translation), each of which solves the Euler-Lagrange equation, and so picking one is the spontaneous breaking of scale-symmetry in the auxiliary problem. Generally we then expect, as happens in our simulation, that $E_{\mathbf{A},\mathbf{Y}}\left(\tilde{\mathbf{v}}\right)=0$ determines a hyper surface in the space of deformation gradients $\mathbf{A}\in\mathbb{R}^{d\times d}$ and that $E_{\mathbf{A},\mathbf{Y}}\left(\tilde{\mathbf{v}}\right)$  changes sign as this hypersurface is crossed. This implies that some form of quasiconvexity is just lost at the critical deformation gradient and corresponds to the onset of metastability in the original problem because the system develops an instability toward nucleation (i.e. the scaling motion of $\tilde{\mathbf{v}}$) while remaining linearly stable. However, we do not know if the nucleation process is itself sub-critical or super-critical with respect to the violation of quasiconvexity, nor can we anticipate the nucleation dynamics.

When the variation of strain (and the domain shape at a boundary point) act to stabilize the nucleation instability, a T-bifurcation results and  this bifurcating branch acts as a center manifold, guiding the evolution of the instability. Because these local variations act to keep the growing nucleus compact, the bifurcation and dynamics will be insensitive to the geometry at large distances from $\mathbf{X}_{0}$. This is {\em not} necessarily true if the variations in strain and geometry are not stabilizing. In this case the nucleation instability becomes sub-critical with respect to the violation of quasiconvexity, and we can expect that the nucleation dynamics will become quite complicated. 

\begin{figure}
\includegraphics[width=.95\linewidth]{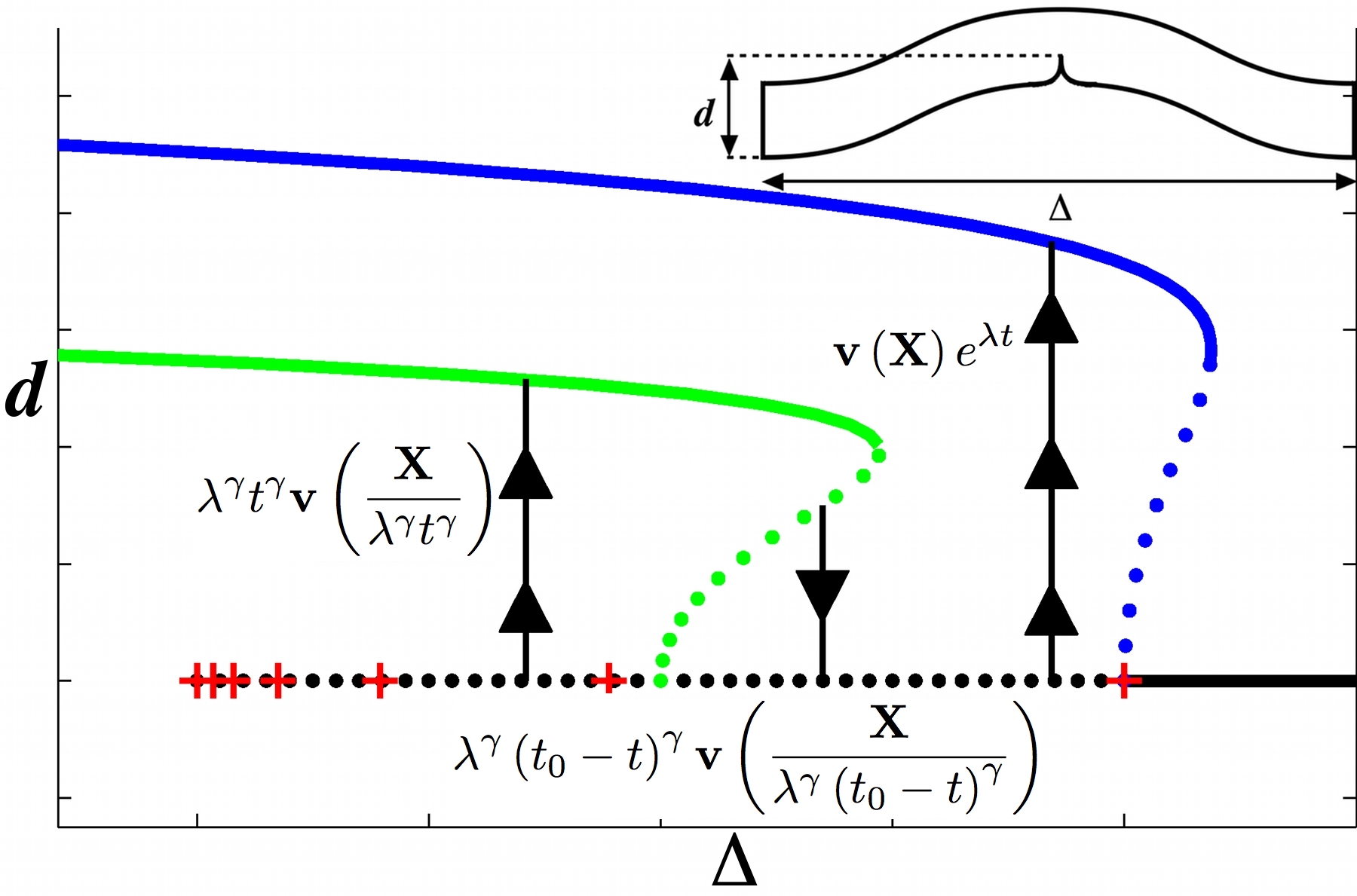}

\caption{Here we show a schematic bifurcation diagram for the finite beam shown in the inset. This geometry destabilizes both the Biot-point and the T-point leading to a hierarchy of sub-critical buckling instabilities accumulating at the Biot-point (red crosses and blue line), and sub-critical sulcification (green line). Black arrow lines schematically show the dynamics  as explained in the text.}
\label{generic}
\end{figure}

As a simple example, consider a uniformly compressed beam of incompressible rubber such as in the inset to Supplementary Fig. \ref{generic}; this figure sketches a hypothetical bifurcation diagram for this system. In this free beam geometry, surface waves on opposite sides of the beam interact to break Biot's surface instability into a hierarchy of sub-critical buckling modes; a qualitative bifurcation diagram for the first buckling mode is traced by the blue line in the figure. Red crosses mark subsequent linear instabilities in the straight beam and tend to an accumulation point marking the Biot point. When the beam is compressed beyond the buckling threshold, a small initial disturbance grows exponentially in time.

Quasiconvexity is still lost at the critical strain of $ \sim35\%$, but could happen before or after the onset of buckling, depending on the aspect ratio of the beam. In this geometry, sulci nucleated at different points can interact with each other and with surface waves in a dynamical
setting. In particular, the inset to Supplementary Fig. \ref{generic}  suggests that buckling and sulcification can interact to enhance one another: sulcification lowers the beams effective bending stiffness while bending increases the local compressive strain which drives sulcification. The resulting sub-critical instability might be a kind of kinked buckling, but the precise location of the kinks, and so the precise development of the instability will be sensitive to imperfections and the initial conditions. In the simplest case when the instability is initiated by a localized perturbation near the midpoint of the beam and the strain in the initial state is above $\sim35\%$, we expect the instability initially resembles a nucleation process governed by the scaling form  {\[\mathbf{x}_{0}\left(\mathbf{X}\right)+L\left(t-t_{0}\right)\mathbf{v}\left(\frac{\mathbf{X}}{L\left(t-t_{0}\right)}\right)\]  with the function $L\left(t\right)$ determined by the dynamical law and the time $t_{0}$ determined by initial conditions. For example if \[\frac{\partial \mathbf{x}}{\partial t}=-\nabla \cdot \frac{\partial W}{\partial \mathbf{A}}\left(\mathbf{\nabla x}\right),\] which describes gradient descent, then {$L\left(t\right)=\lambda^{\gamma}t^{\gamma}$ with} $\gamma=1/2$. If \[\frac{\partial^{2} \mathbf{x}}{\partial t^{2}}=\nabla \cdot \frac{\partial W}{\partial \mathbf{A}}\left(\mathbf{\nabla x}\right),\] which describes inertial dynamics, then $L\left(t\right)$ has the same form with $\gamma=1$. If the dynamics are viscoelastic, but local in time, e.g. if the dynamical law has the form \[\nabla \cdot \mathbf{S}\left(\mathbf{\nabla x}, \mathbf{\nabla} \frac{\partial \mathbf{x}}{\partial t}\right)  =0,\] where $\mathbf{S}$  is the nominal strain, then  $L\left(t\right)=e^{t/\lambda}$. The green curve traces the bifurcation diagram for this process. Since kinked buckling is sub-critical, below the critical strain where quasiconvexity is lost, a small size sulcus will shrink according to the scaling dynamical law: \[ \mathbf{x}_{0}\left(\mathbf{X}\right)+L\left(t_{0}-t\right)\mathbf{v}\left(\frac{\mathbf{X}}{L\left(t_{0}-t\right)}\right).\]

Notice that when the dynamics are either inertial or gradient descent, the sulcus will nucleate (or vanish) at a finite time, but that for viscoelastic material, the sulcus cannot form spontaneously (or completely vanish). Assuming that PDMS is governed by a viscoelastic dynamics, this potentially explains why the metastable portion of the computed bifurcation diagram is experimentally accessible and the existence of the vertical black line in SI Fig. 2(c)(4).}

\bibliographystyle{plain}
\bibliography{creasing}


\end{document}